\documentclass[pdflatex,sn-nature]{sn-jnl}

\usepackage{graphicx}%
\usepackage{multirow}%
\usepackage{amsmath,amssymb,amsfonts}%
\usepackage{amsthm}%
\usepackage{mathrsfs}%
\usepackage[title]{appendix}%
\usepackage{xcolor}%
\usepackage{textcomp}%
\usepackage{manyfoot}%
\usepackage{booktabs}%
\usepackage{algorithm}%
\usepackage{algorithmicx}%
\usepackage{algpseudocode}%
\usepackage{listings}%

\theoremstyle{thmstyleone}%
\theoremstyle{thmstyletwo}%

\theoremstyle{thmstylethree}%

\begin{document}

\title[Article Title]{Personalized targeted memory reactivation enhances consolidation of challenging memories via slow wave and spindle dynamics}


\author[1]{\fnm{Gi-Hwan} \sur{Shin}}\email{gh\_shin@korea.ac.kr}

\author[1]{\fnm{Young-Seok} \sur{Kweon}}\email{youngseokkweon@korea.ac.kr}

\author[2]{\fnm{Seungwon} \sur{Oh}}\email{swoh@kongju.ac.kr}

\author*[3]{\fnm{Seong-Whan} \sur{Lee}}\email{sw.lee@korea.ac.kr}

\affil[1]{\orgdiv{Department of Brain and Cognitive Engineering}, \orgname{Korea University}, \orgaddress{\city{Seoul}, \country{Republic of Korea}}}

\affil[2]{\orgdiv{Department of Artificial Intelligence}, \orgname{Kongju National University}, \orgaddress{\city{Cheonan}, \country{Republic of Korea}}}

\affil[3]{\orgdiv{Department of Artificial Intelligence}, \orgname{Korea University}, \orgaddress{\city{Seoul}, \country{Republic of Korea}}}


\abstract{Sleep is crucial for memory consolidation, underpinning effective learning. Targeted memory reactivation (TMR) can strengthen neural representations by re-engaging learning circuits during sleep. However, TMR protocols overlook individual differences in learning capacity and memory trace strength, limiting efficacy for difficult-to-recall memories. Here, we present a personalized TMR protocol that adjusts stimulation frequency based on individual retrieval performance and task difficulty during a word-pair memory task. In an experiment comparing personalized TMR, TMR, and control groups, the personalized protocol significantly reduced memory decay and improved error correction under challenging recall. Electroencephalogram (EEG) analyses revealed enhanced synchronization of slow waves and spindles, with a significant positive correlation between behavioral and EEG features for challenging memories. Multivariate classification identified distinct neural signatures linked to the personalized approach, highlighting its ability to target memory-specific circuits. These findings provide novel insights into sleep-dependent memory consolidation and support personalized TMR interventions to optimize learning outcomes.}

\maketitle

\section{INTRODUCTION}
Sleep plays a fundamental role in memory consolidation—a process critical for effective learning and educational achievement \cite{abdellahi2023targeting}. Beyond mere memory preservation, sleep actively transforms newly encoded information into long-term representations through reorganization and selective stabilization processes \cite{diekelmann2010memory}. These processes are thought to underlie the system consolidation framework, whereby memories are gradually transferred from hippocampal to neocortical storage sites. Converging evidence suggests that sleep-specific brain oscillations support this transformation by replaying and enhancing memory traces \cite{stickgold2006memory}. Although both rapid eye movement (REM) and non-REM (NREM) sleep contribute to memory processing \cite{lendner2023human, li2017rem, carbone2024comparing, rudoy2009strengthening, cairney2018memory}, extensive research has focused on NREM sleep—particularly during stages NREM 2 and NREM 3—because it is associated with robust oscillatory events that stabilize and integrate newly acquired information \cite{whitmore2024targeted}. Electroencephalogram (EEG) recordings during NREM sleep consistently reveal two key oscillatory dynamics—slow waves (SW) and sleep spindles—that are integral to this consolidation process \cite{bar2020local, chen2024modulating}. Specifically, SW activity promotes widespread cortical synchrony and maintains synaptic homeostasis, priming the brain for efficient information processing \cite{rasch2007odor, marshall2006boosting, schreiner2021endogenous}. Concurrently, sleep spindles, generated by thalamocortical circuits, enhance hippocampal-cortical communication and facilitate robust memory consolidation \cite{buzsaki1996hippocampo}. Together, these oscillations synchronize neural activity across distributed brain regions, reinforcing memory reactivation and integration into established cortical networks \cite{daume2024control, staresina2015hierarchical}. Despite considerable advances in our understanding of these intrinsic processes, a key challenge remains: developing external interventions that can effectively synchronize with and further enhance these oscillatory mechanisms to optimize memory consolidation and support improved learning outcomes.

Targeted memory reactivation (TMR) has emerged as a promising sleep-based intervention to enhance memory consolidation by delivering task-relevant cues during sleep \cite{oudiette2013upgrading, whitmore2022targeted}. By aligning with the natural synchronization of SW and spindle activity, TMR enhances cortical plasticity and facilitates large-scale reorganization across brain networks \cite{hu2020promoting, carbone2024update}. A growing number of studies have reported improvements in various memory tasks—including word recall, spatial recognition, and motor skill learning \cite{wang2019targeted, abdellahi2023targeted, goldi2019effects}—which has sparked interest in the potential of TMR as an intervention to enhance learning outcomes and cognitive performance in both educational \cite{vidal2022odor, neumann2020odor} and clinical settings \cite{yuan2021effect}. However, despite its promise, optimal implementation of TMR remains under debate. Fixed cueing protocols may yield inconsistent outcomes by failing to account for individual variability in learning capacity and memory trace strength \cite{creery2015targeted}.

Recent research further indicates that TMR efficacy is influenced not only by cueing during sleep but also by the quality of pre-sleep learning \cite{antony2017retrieval, liu2023item}. Studies demonstrate that TMR is particularly effective for recalling familiar or moderately strong memories, whereas its benefits diminish for unfamiliar or overly complex items \cite{klaassen2023difficulty}. In combination with models proposing an inverted U-shaped relationship between memory strength and consolidation \cite{bauml2014sleep, stickgold2009remember}, these findings reveal that fixed TMR protocols face significant limitations, especially for challenging recall tasks. This evidence underscores the necessity for personalized approaches that customize cueing strategies to individual differences and task demands to optimize memory reactivation.

\begin{figure}[t!]
\centering
\includegraphics[width=1.00\textwidth]{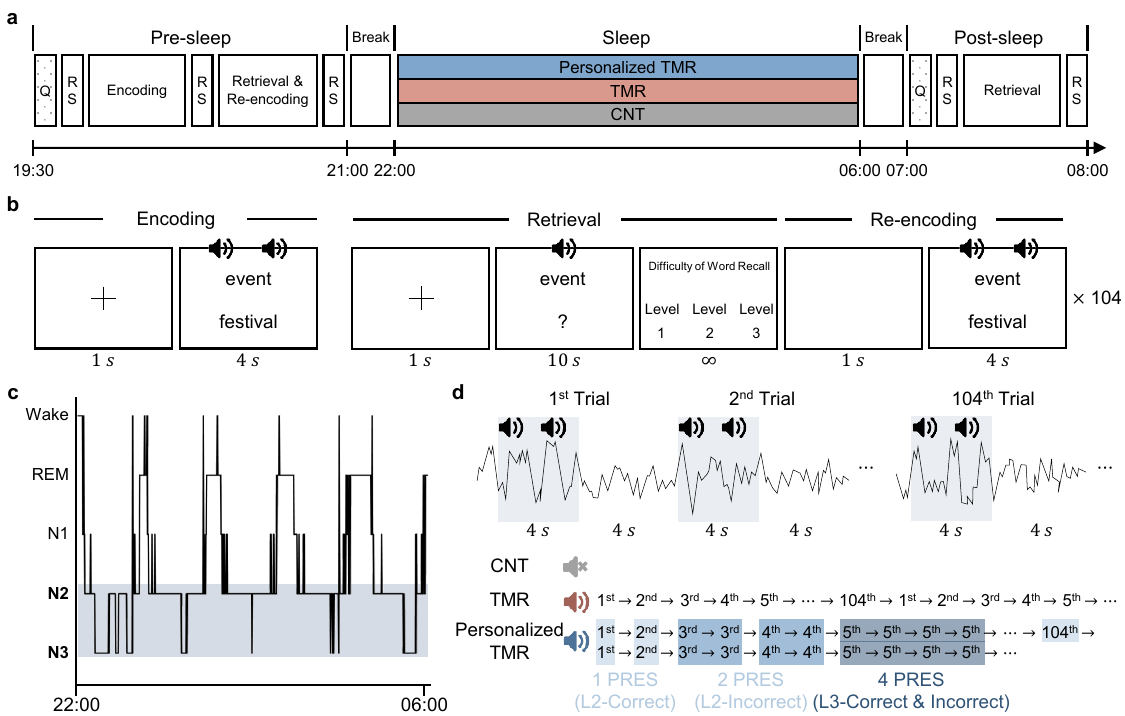}
\caption{Experimental setup. \textbf{a} The experiment comprises three main sessions: pre-sleep, sleep, and post-sleep. Participants were randomly assigned to one of three groups during the sleep session: the personalized TMR, TMR, and CNT groups. The questionnaire (Q) was administered during the wakefulness periods, alongside tasks such as word-pair learning and a 5-min eyes-closed resting state (RS) in both pre- and post-sleep sessions. \textbf{b} In the encoding phase of the pre-sleep session, participants were exposed to 104-word pairs presented via audiovisual stimuli. In the retrieval phase (pre- and post-sleep sessions), participants recalled and typed the associated word upon viewing the learned cue. Immediately following each retrieval trial, participants rated their expected difficulty in recalling the word pair both immediately and after a 12-h delay using a three-level scale: L1, L2, and L3. These prospective self-ratings were used to classify word-pair difficulty levels and were not updated or redefined across retrieval rounds. Re-encoding was provided only during the pre-sleep retrieval phase. \textbf{c} A hypnogram displaying the sleep session for one participant shows the NREM 2 and NREM 3 stages, during which the stimuli were administered, indicated by gray shading. \textbf{d} Stimulation protocols varied among groups: the personalized TMR group adjusted the number of presentations (PRES) based on response correctness—1 PRES for L2 correct responses, 2 PRES for L2 incorrect responses, and 4 PRES for L3 correct and incorrect responses. The TMR group received the same stimuli uniformly, while the CNT group received no stimulation.}
\end{figure}

To address these challenges, we developed a personalized TMR protocol that tailors auditory stimulation frequency based on individual recall ability and task difficulty (Fig. 1). To evaluate this personalized approach, we designed an experiment with thirty-six participants (13 females, mean age: 26.57 ± 3.13 years) who performed a memory task across two sessions—pre-sleep and post-sleep. In the pre-sleep session, 104 related word pairs were presented via audiovisual stimuli during encoding. During retrieval in both sessions, participants recalled and typed the associated word upon cue presentation and rated recall difficulty on three levels (L1, L2, or L3); re-encoding was provided in the pre-sleep session regardless of response accuracy. In the sleep session, participants were randomly assigned to one of three groups, with each group comprising 12 individuals: the personalized TMR group received auditory cues tailored to recall difficulty (with 1, 2, or 4 presentations (PRES) based on performance), the TMR group received uniform stimulation for all word pairs, and the control (CNT) group received no auditory stimulation. The CNT group was included as a no-stimulation baseline condition to isolate the effects of cueing frequency and personalization on memory consolidation, rather than to assess the general impact of auditory stimulation during sleep. Auditory cues were delivered during stable NREM 2 and NREM 3 stages, with stimulation paused if wake, REM, or NREM 1 stages were detected (see Methods).

Building on our experimental design, we hypothesize that personalized auditory stimulation—tailored to individual retrieval performance and task difficulty—will lead to more robust memory consolidation than fixed cueing protocols. Specifically, we predict that personalized TMR will more effectively reduce memory decay and enhance error correction, particularly for challenging memories such as word pairs that are more difficult to recall. To test this hypothesis, we compared personalized TMR, TMR, and CNT groups, assessing both behavioral performance and neural dynamics using EEG. Our planned analyses include correlational assessments between behavioral improvements and EEG features as well as multivariate classification to identify distinct neural signatures associated with the personalized intervention. Overall, our study aims to advance our understanding of sleep-dependent memory consolidation and to establish a foundation for personalized, sleep-based cognitive enhancement strategies with implications for learning and education.

\section{RESULTS}
\subsection{Sleep-dependent memory performance across recall difficulty levels}
We compared demographics, questionnaires, and sleep architecture across the three groups (Supplementary Table 1). No significant group differences were found, confirming that memory performance changes were driven by intervention strategies rather than variations in participant characteristics or sleep parameters.

To validate our prospective difficulty classification, we examined the association between subjective difficulty ratings and retrieval accuracy. Strong negative correlations were consistently observed across all groups and sessions (\textit{rho} = –0.878 to –0.948; all \textit{p} $<$ 0.001), indicating that higher difficulty ratings were reliably associated with lower recall performance. Memory accuracy across recall difficulty levels (All, L1, L2, and L3) was analyzed using two-way analysis of variance (ANOVA) to assess the effects of group (personalized TMR, TMR, and CNT) and time (pre- and post-sleep) (Fig. 2, top row). Bonferroni correction was applied to all post-hoc analyses. In the All condition, significant main effects of time (\textit{F}$_{1,66}$ = 42.42, \textit{p} $<$ 0.001) and group (\textit{F}$_{2,66}$ = 5.10, \textit{p} = 0.009) were observed, but no significant interaction effect (\textit{F}$_{2,66}$ = 0.74, \textit{p} = 0.479) was found. Paired \textit{t}-tests showed significant pre- to post-sleep improvements in all groups: personalized TMR (\textit{t}$_{11}$ = -10.22, \textit{p} $<$ 0.001), TMR (\textit{t}$_{11}$ = -8.37, \textit{p} $<$ 0.001), and CNT (\textit{t}$_{11}$ = -6.93, \textit{p} $<$ 0.001). In contrast, post-hoc comparisons showed no significant group differences in either session. For the L1 condition, a significant group effect was observed (\textit{F}$_{2,66}$ = 3.61, \textit{p} = 0.033), while effects of time (\textit{F}$_{1,66}$ = 1.85, \textit{p} = 0.179) and interaction (\textit{F}$_{2,66}$ = 0.63, \textit{p} = 0.538) were not significant. Post-hoc paired \textit{t}-tests revealed no significant changes from pre- to post-sleep in any group. In the L2 condition, significant effects of group (\textit{F}$_{2,66}$ = 9.47, \textit{p} $<$ 0.001) and time (\textit{F}$_{1,66}$ = 12.55, \textit{p} $<$ 0.001) were observed, while the interaction effect was not (\textit{F}$_{2,66}$ = 1.32, \textit{p} = 0.274). Paired \textit{t}-tests revealed significant improvements in personalized TMR (\textit{t}$_{11}$ = -3.73, \textit{p} = 0.003) and TMR (\textit{t}$_{11}$ = -5.65, \textit{p} = 0.001), but not in CNT (\textit{t}$_{11}$ = -1.41, \textit{p} = 0.187). personalized TMR outperformed CNT in the post-sleep session (\textit{t}$_{22}$ = 4.33, \textit{p} $<$ 0.001). In the most demanding L3 condition, significant effects of group (\textit{F}$_{2,66}$ = 16.54, \textit{p} $<$ 0.001), time (\textit{F}$_{1,66}$ = 20.77, \textit{p} $<$ 0.001), and their interaction (\textit{F}$_{2,66}$ = 8.91, \textit{p} $<$ 0.001) were observed. Paired \textit{t}-tests showed improvements in personalized TMR (\textit{t}$_{11}$ = -5.68, \textit{p} $<$ 0.001), but not in TMR (\textit{t}$_{11}$ = -1.50, \textit{p} = 0.161) or CNT (\textit{t}$_{11}$ = -1.56, \textit{p} = 0.148). personalized TMR performed significantly better than both TMR (\textit{t}$_{22}$ = 3.96, \textit{p} $<$ 0.001) and CNT (\textit{t}$_{22}$ = 4.49, \textit{p} $<$ 0.001) in the post-sleep session. For differences in accuracy between the pre-sleep and post-sleep sessions among groups (Fig. 2, bottom row), one-way ANOVA revealed significant differences in the All (\textit{F}$_{2,33}$ = 3.83, \textit{p} = 0.032), L2 (\textit{F}$_{2,33}$ = 4.00, \textit{p} = 0.028), and L3 (\textit{F}$_{2,33}$ = 14.18, \textit{p} $<$ 0.001) conditions. Post-hoc comparisons showed that personalized TMR outperformed CNT in the All condition (\textit{t}$_{22}$ = 2.69, \textit{p} = 0.013), while TMR outperformed CNT in the L2 condition (\textit{t}$_{22}$ = 2.94, \textit{p} = 0.008). In the L3 condition, personalized TMR outperformed both TMR (\textit{t}$_{22}$ = 3.89, \textit{p} = 0.008) and CNT (\textit{t}$_{22}$ = 4.54, \textit{p} = 0.002).

\begin{figure}[t!]
\centering
\includegraphics[width=1.00\textwidth]{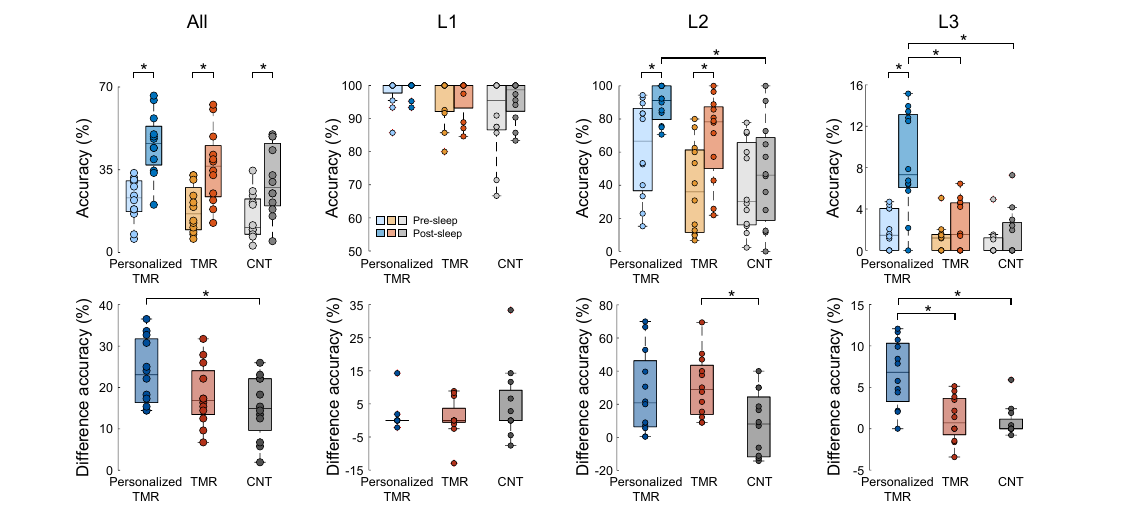}
\caption{Memory accuracy. Memory accuracy for all trials and for each difficulty level (L1, L2, and L3) is shown for the personalized TMR, TMR, and CNT groups. In the top row, the light-colored bars represent pre-sleep session accuracy and the dark-colored bars represent post-sleep session accuracy. The bottom row displays the difference in accuracy between the pre-sleep and post-sleep sessions. Significant differences (two-sample \textit{t}-tests with Bonferroni correction, \textit{p} $<$ 0.05) are marked with asterisks.}
\end{figure}

Memory transitions from pre-sleep to post-sleep sessions were analyzed as ratios by recall level and categorized into four types: correct-correct (CC), correct-incorrect (CI), incorrect-correct (IC), and incorrect-incorrect (II) (Fig. 3, top row). As a result, all groups showed generally higher ratios for II transition, with a significantly lower proportion of CI transition. One-way ANOVA with Bonferroni correction was used for group comparisons (Fig. 3, bottom row). No significant group differences were found for CC and CI transitions, indicating comparable retention and incorrect recall rates across groups and conditions. Significant group differences were observed for IC transition in both the All (\textit{F}$_{2,33}$ = 4.13, \textit{p} = 0.025) and L3 (\textit{F}$_{2,33}$ = 3.43, \textit{p} = 0.044) conditions. Post-hoc comparisons revealed that personalized TMR had significantly higher IC ratios compared to CNT in both the All (\textit{t}$_{22}$ = 2.84, \textit{p} = 0.010) and L3 (\textit{t}$_{22}$ = 2.63, \textit{p} = 0.015) conditions, suggesting effective error correction in personalized TMR. For II transition, significant group differences were found in the All condition (\textit{F}$_{2,33}$ = 3.45, \textit{p} = 0.044). Post-hoc comparisons revealed that CNT exhibited higher II ratios than personalized TMR (\textit{t}$_{22}$ = -2.70, \textit{p} = 0.013), reflecting greater persistence of incorrect responses in CNT.

\begin{figure}[t!]
\centering
\includegraphics[width=1.00\textwidth]{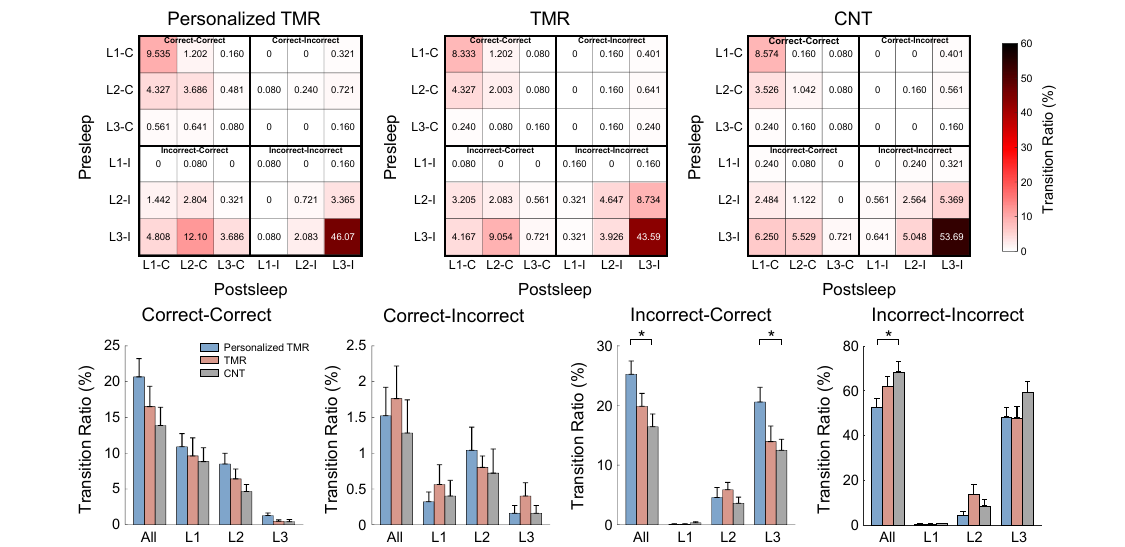}
\caption{Memory transition. Memory transition ratios for four outcomes (correct-correct, correct-incorrect, incorrect-correct, and incorrect-incorrect) are presented for the personalized TMR, TMR, and CNT groups. The top row shows a heatmap of transition ratios, and the bottom row presents a bar plot of mean transition values and standard errors, calculated across participants and recall levels. Significant differences (two-sample \textit{t}-tests with Bonferroni correction, \textit{p} $<$ 0.05) are indicated with asterisks.}
\end{figure}

\subsection{Dynamic neural responses to auditory stimulation during sleep}
To investigate neural mechanisms underlying memory consolidation during sleep, we analyzed event-related potential (ERP), time-frequency representation (TFR), and event-related phase-amplitude coupling (ERPAC) in response to auditory stimulation across three groups (Fig. 4). Statistical analyses were performed using one-way ANOVA followed by Bonferroni-corrected post-hoc \textit{t}-tests to compare neural responses among these groups.

ERP responses revealed distinct waveform changes in the personalized TMR and TMR groups compared to the CNT group (Fig. 4a). In the All condition, the personalized TMR group exhibited significantly negative amplitudes immediately after the second stimulus, followed by positive amplitudes during the 3–4 s interval. The TMR group displayed negative amplitudes during the same interval. These distinct temporal dynamics highlight measurable differences in the processing of auditory stimuli between the personalized TMR and TMR groups. In the L3 condition, similar patterns emerged. The personalized TMR group showed significantly greater neural responsiveness than both the TMR and CNT groups, indicating heightened neural activity under increased recall difficulty. To further examine whether ERP responses varied across repeated cue presentations, we compared the first and last cue within each 4-PRES sequence in the personalized TMR group. This analysis revealed significant amplitude differences, suggesting neural adaptation across repetitions during sleep (Supplementary Fig. 1). Additionally, channel-wise ERP analysis revealed that significant group differences were primarily observed in frontal (F3 and F4) and central (C3 and C4) regions, whereas occipital channels (O1 and O2) showed minimal modulation (Supplementary Fig. 2). This spatial pattern indicates that cue-evoked neural responses during sleep were most prominently expressed in anterior cortical regions.

TFR analyses showed enhanced activation within the SW and spindle bands in both the personalized TMR and TMR groups compared to the CNT group (Fig. 4b). During the initial 0–0.5 s post-stimulus interval, the TMR group exhibited significantly greater SW activation compared to personalized TMR. In the 2–4 s post-stimulus interval, the personalized TMR group demonstrated pronounced increases in both SW and spindle power compared to CNT, reflecting sustained neural activation. A similar pattern was observed in the TMR group, which showed consistent increases in SW and spindle activation relative to CNT across the entire post-stimulus period. These distinct features highlight group-specific spectral dynamics. Channel-wise TFR analysis further revealed that these group differences were most pronounced over frontal and central regions, whereas occipital channels showed relatively attenuated spectral responses (Supplementary Fig. 3), suggesting regionally specific engagement of memory-related oscillatory activity.

To quantify cross-frequency interactions associated with memory reactivation, we computed ERPAC across a broad frequency range (4–20 Hz), focusing on SW phase and spindle-band amplitude interactions (Fig. 4c). Pronounced coupling was observed within the spindle band, reflecting enhanced SW-spindle dynamics. In the All condition, both the personalized TMR and TMR groups exhibited higher ERPAC values compared to the CNT group, indicating stronger cross-frequency coupling elicited by auditory stimulation. In the L3 condition, ERPAC values were highest in the TMR group, indicating robust cross-frequency interactions under heightened recall difficulty. In the personalized TMR group, ERPAC values were slightly lower, potentially reflecting auditory habituation effects caused by repeated stimulation. These findings highlight distinct neural dynamics between personalized TMR and TMR under challenging recall conditions.

\begin{figure}[t!]
\centering
\includegraphics[width=1.00\textwidth]{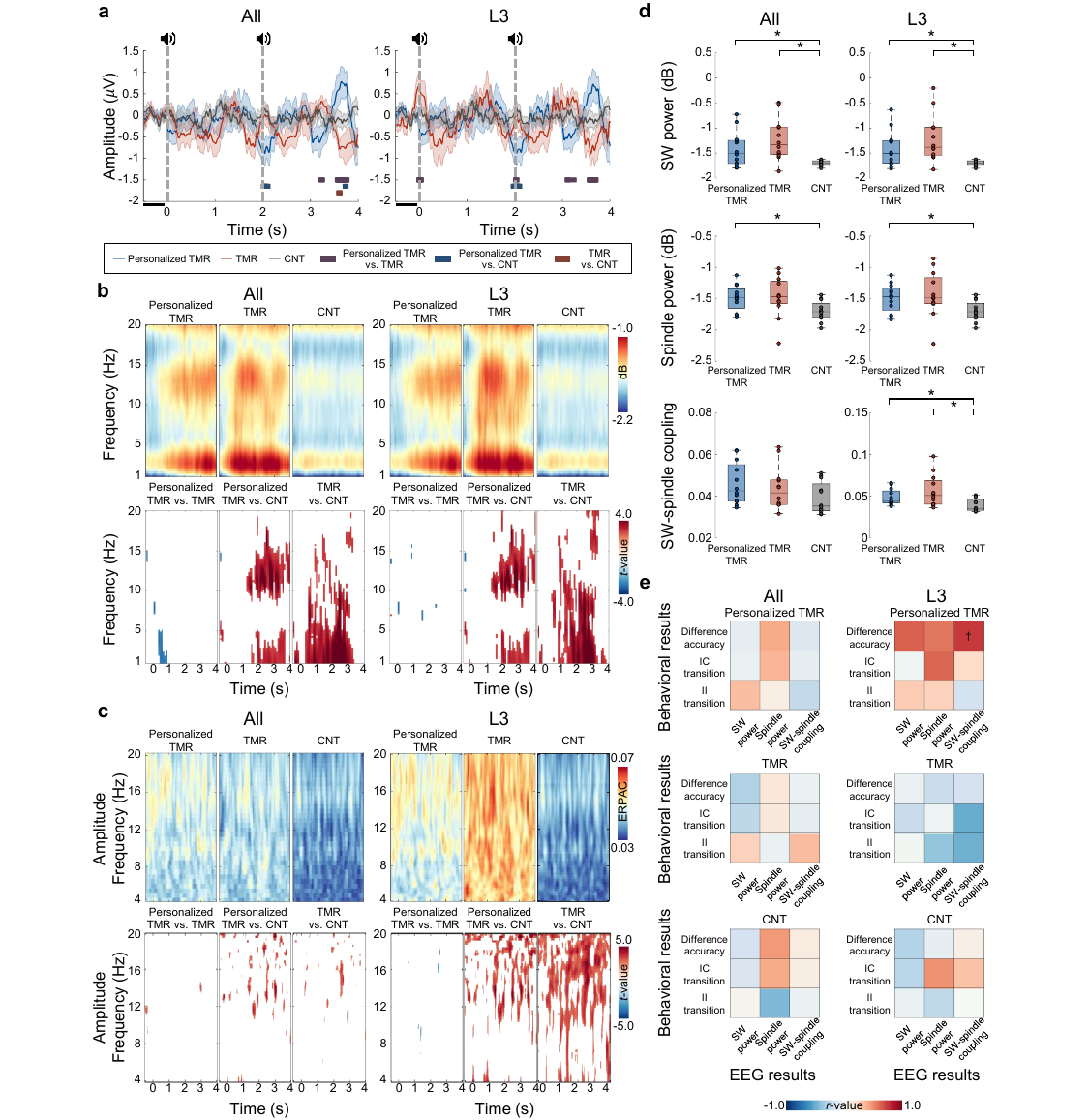}
\caption{Electrophysiological responses to auditory stimuli during sleep. \textbf{a} Average event-related potentials (ERP) are displayed across all channels for the three groups in response to auditory cues. Each auditory cue consisted of a word pair presented sequentially within a 4-s window: the first word was delivered at stimulus onset (0 s), followed by the second word approximately 2 s later. The baseline period (-0.5 to 0 s) is indicated by a black bar, and stimulus onset is marked by a dashed line. Colored lines indicate group-specific ERP waveforms, and horizontal bars above the waveforms denote time intervals showing significant group differences. \textbf{b} Time-frequency representations are averaged across channels, with the top row displaying group-level spectral responses and the bottom row highlighting significant intergroup differences. \textbf{c} Event-related phase-amplitude coupling results illustrate slow wave (SW)-spindle coupling, with the top row presenting group results and the bottom row indicating significant differences. \textbf{d} Comparisons of SW power, spindle power, and SW–spindle coupling across groups reveal significant differences; significant findings from two-sample \textit{t}-tests with Bonferroni correction (p $<$ 0.05) are marked with asterisks. \textbf{e} Pearson’s correlations between behavioral and EEG results indicate significant interactions, marked with a dagger (\textit{p} $<$ 0.05, FDR correction).}
\end{figure}

\subsection{Comparison on key EEG features across groups}
We compared key EEG features (SW power, spindle power, and SW-spindle coupling) across groups (Fig. 4d). In the All condition, significant group differences in SW power were observed (\textit{F}$_{2,33}$ = 6.52, \textit{p} = 0.004). Post-hoc \textit{t}-tests indicated that SW power was significantly higher in the personalized TMR group compared to CNT (\textit{t}$_{22}$ = 2.72, \textit{p} = 0.013) and in the TMR group compared to CNT (\textit{t}$_{22}$ = 3.79, \textit{p} = 0.001). In the L3 condition, similar group differences were found (\textit{F}$_{2,33}$ = 6.01, \textit{p} = 0.006). Post-hoc analyses revealed significantly higher SW power in the personalized TMR group compared to CNT (\textit{t}$_{22}$ = 2.69, \textit{p} = 0.013) and in the TMR group compared to CNT (\textit{t}$_{22}$ = 3.53, \textit{p} = 0.002). For spindle power, group differences were significant in the All condition (\textit{F}$_{2,33}$ = 3.31, \textit{p} = 0.049) and the L3 condition (\textit{F}$_{2,33}$ = 3.34, \textit{p} = 0.048). Post-hoc \textit{t}-tests showed that spindle power was significantly higher in the personalized TMR group compared to CNT in both the All (\textit{t}$_{22}$ = 2.77, \textit{p} = 0.011) and L3 (\textit{t}$_{22}$ = 2.63, \textit{p} = 0.015) conditions. For SW-spindle coupling, significant group differences were observed exclusively in the L3 condition (\textit{F}$_{2,33}$ = 5.02, \textit{p} = 0.013). Post-hoc analyses revealed that both the personalized TMR (\textit{t}$_{22}$ = 2.62, \textit{p} = 0.016) and TMR (\textit{t}$_{22}$ = 2.86, \textit{p} = 0.009) groups exhibited stronger coupling compared to CNT. These findings demonstrate distinct neural mechanisms engaged by personalized TMR and TMR, highlighting their differential contributions to memory-related neural dynamics during sleep.

\subsection{Behavioral and neural correlates of memory consolidation}
We examined how neural dynamics relate to memory consolidation by correlating behavioral and EEG results (Fig. 4e, Supplementary Fig. 4). In the L3 condition, a significant correlation was observed exclusively in the personalized TMR group: differences in accuracy showed a strong positive association with SW-spindle coupling (\textit{r} = 0.70, \textit{p} = 0.011). This finding highlights the importance of cross-frequency interactions in supporting memory reactivation and stabilization. By contrast, correlations involving SW power and spindle power were not statistically significant. No significant correlations were found in the All condition across any group, underscoring the relevance of enhanced neural dynamics in the personalized TMR group under heightened recall difficulty. These findings further support the distinct contributions of personalized reactivation strategies to memory consolidation, as evidenced by the significant correlation between memory performance and SW–spindle coupling in the personalized TMR group.

\begin{figure}[t!]
\centering
\includegraphics[width=0.60\textwidth]{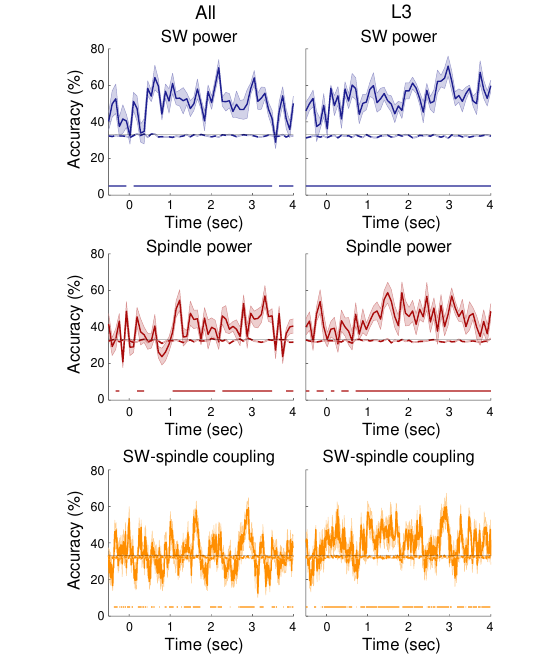}
\caption{Classification results. Multivariate classification performance for distinguishing personalized TMR, TMR, and CNT groups in All and level 3 (L3) conditions based on EEG-derived features (slow wave (SW) power, spindle power, and SW–spindle coupling) is presented. The shaded solid line represents observed decoding performance, while the shaded dashed line shows surrogate decoding results from 250 label shuffles. Accuracy values are presented with means and standard errors. The horizontal black solid line indicates chance-level performance, and the lower horizontal line highlights statistically significant time points, identified via a two-sided cluster-based permutation test with 1,000 randomizations.}
\end{figure}

\subsection{Classification of EEG features revealing neural dynamics across groups}
We investigated memory dynamics during sleep by performing multivariate classification of EEG features across groups (Fig. 5). Each EEG feature—SW power, spindle power, and SW-spindle coupling—was independently classified using an SVM with an RBF kernel, revealing significant group distinctions. The accuracy values confirmed the strong discriminative power of these features across all conditions, with classification performance exceeding chance levels. Notably, accuracy values were higher after the second stimulus (2 s) compared to the first (0 s), peaking at around 3 s. This peak suggests robust neural dynamics associated with memory processing during this interval. Similarly, AUC results demonstrated consistent performance across conditions, further corroborating the robustness of these findings (Supplementary Fig. 5). Surrogate decoding analysis verified that classification performance significantly exceeded chance levels, highlighting the sensitivity of EEG features to memory reactivation dynamics during sleep.

\section{DISCUSSION}
This study highlights the unique efficacy of personalized TMR, a personalized protocol that aligns with individual learning capacities, in enhancing memory consolidation by precisely modulating SW and spindle dynamics. Behavioral analysis revealed pronounced effects in the most challenging condition, where task complexity maximized sleep-dependent reactivation. During NREM sleep, improvements induced by personalized TMR were supported by significant modulations in SW power, spindle power, and their coupling, highlighting the ability of the protocol to fine-tune neural reactivation processes. Finally, multivariate analyses revealed distinct neural signatures specific to the personalized TMR group, emphasizing its differentiation from conventional fixed-stimulation protocols.

The behavioral results underscore the critical role of tailoring stimulation to individual learning capacities for effective memory consolidation. In the All condition, the personalized TMR group consistently outperformed other groups in memory improvement. The absence of significant changes in the L1 condition likely reflects the simplicity of tasks that were already consolidated before sleep, reducing reliance on nocturnal reactivation \cite{bauml2014sleep, cairney2016benefits}. This suggests that targeted stimulation offers limited advantages for tasks with minimal dependency on sleep-based reactivation. In contrast, the L2 condition showed improvements in both personalized TMR and TMR groups, supporting the efficacy of targeted stimulation for tasks of moderate complexity, as observed in prior studies \cite{stickgold2009remember}. However, significant accuracy differences were noted only in the TMR group compared to the CNT group, potentially due to the limited number of stimuli presented in the personalized TMR protocol \cite{ngo2013auditory}. The most pronounced effects emerged in the L3 condition, where the personalized TMR group significantly outperformed both the TMR and CNT groups, demonstrating its ability to dynamically adjust to task complexity and enhance memory reactivation under heightened cognitive demands \cite{bendor2012biasing}. Additionally, the personalized TMR group exhibited significantly higher IC transition, suggesting enhanced reinforcement of weaker memory traces through precise, task-aligned reactivation \cite{brodt2018fast}. Conversely, reduced II transition reflect the ability of personalized TMR to minimize the retention of unstable memory traces, highlighting its role in error correction and memory stabilization \cite{schreiner2015boosting}.

The findings reveal distinct neural mechanisms underpinning memory consolidation facilitated by auditory stimulation during sleep. Changes in ERP amplitudes observed in the personalized TMR and TMR groups compared to the CNT group indicate heightened neural responsiveness, which may reflect memory-related processing during sleep \cite{cairney2018memory}, but could also result from non-specific sensory responses to auditory input \cite{atienza2001auditory}. In the personalized TMR group, pronounced negative amplitudes following the second stimulus suggest robust sensory gating within the thalamocortical network, prioritizing task-relevant stimuli and optimizing neural resources for memory reactivation \cite{wang2010thalamic, jones2016cognitive}. During the 3–4 s interval, ERP amplitudes revealed distinct group differences. The personalized TMR group showed stronger cue-evoked ERP responses than the TMR group, a pattern consistent with prior findings suggesting that such responses may reflect memory-related information processing during sleep \cite{cairney2018memory}. This enhanced responsiveness may reflect a cumulative effect of repeated cueing within the personalized TMR protocol, which could have progressively shaped neural engagement with memory-relevant stimuli during sleep. Importantly, ERP group differences were most pronounced in frontal and central channels, with minimal effects observed in occipital regions. This topographical pattern aligns with prior findings suggesting that anterior cortical regions are preferentially involved in sleep-related memory processing and cue reactivity \cite{schreiner2015boosting}.

TFR analyses revealed that both the personalized TMR and TMR groups exhibited significant increases in SW and spindle power compared to the CNT group, indicating that the temporal structure of cue delivery may have contributed to overall enhancements in memory-related neural activity. Specifically, each word pair was presented within a 4-s window, with the second word typically ending around 2.9 s after onset. This was followed by a fixed 4-s inter-stimulus interval, resulting in an effective interval of approximately 5 s between the end of one cue and the onset of the next. This interval exceeds the spindle refractory period \cite{antony2019sleep}, during which memory reactivation is less likely to succeed. By aligning with these neurophysiological constraints, our stimulation protocol likely minimized interference between cues and supported more effective consolidation processes \cite{cairney2018memory, farthouat2017new}. Channel-wise TFR analyses further revealed that these spectral enhancements were most prominent in frontal and central regions, with weaker modulation in occipital areas. This spatial distribution is consistent with prior findings demonstrating that SW and sleep spindles are preferentially expressed in anterior cortical areas, particularly during memory-relevant NREM dynamics \cite{molle2002grouping}. Statistical analyses revealed significant increases in SW power in both personalized TMR and TMR groups compared to CNT, reflecting enhanced thalamocortical synchronization critical for memory stabilization during NREM sleep \cite{buzsaki1996hippocampo, levenstein2019nrem}. However, spindle power exhibited a significant increase only in the personalized TMR group, emphasizing its unique role in promoting synaptic plasticity and facilitating hippocampal-cortical communication for memory updating \cite{creery2022electrophysiological}. While repeated stimulation in the personalized protocol optimized SW and spindle activity, it may have also induced stimulus-specific adaptation in the auditory cortex, potentially reducing EEG power relative to the TMR group \cite{ulanovsky2003processing}. These results highlight how personalized TMR optimizes neural oscillations for efficient memory consolidation, even under varying cognitive demands.

Importantly, while previous studies have reported that presenting full cue-feedback pairs during sleep can diminish or block memory reactivation effects \cite{schreiner2015auditory, forcato2020reactivation}, our findings suggest that this outcome may depend on specific boundary conditions. In our study, each word pair was delivered with sufficient spacing and during stable NREM sleep stages, which are known to promote hippocampal–neocortical communication. Moreover, ERP and TFR analyses revealed enhanced neural responses following the second word, particularly in difficult items, indicating that the second word may have functioned as a reinstatement signal that facilitated rather than disrupted reactivation. These results suggest that the effectiveness of full-pair cues may depend on temporal precision, memory strength, and sleep-stage alignment, and that under optimal conditions, full cues can successfully promote memory consolidation.

ERPAC analysis highlights the critical role of SW-spindle coupling in coordinating hippocampal-cortical communication during memory consolidation. Enhanced coupling in the personalized TMR and TMR groups, compared to the CNT group, reflects improved coordination between thalamocortical and hippocampal networks—a fundamental mechanism for stabilizing labile memory traces into durable long-term representations \cite{nicolas2022sigma, schreiner2023respiration, niknazar2015coupling}. In the L3 condition, both personalized TMR and TMR groups demonstrated significantly stronger SW-spindle coupling relative to the CNT group, highlighting the importance of synchronized neural oscillations in meeting heightened recall demands \cite{denis2021sleep}. This synchronization likely enhances neural alignment, reduces interference, and ensures functional specificity for complex memory reactivation \cite{van2012memory,kim2017enhancing}. Conversely, the minimal activation observed in the CNT group underscores the critical role of external stimulation in driving effective memory reactivation \cite{santamaria2024effects}. These findings reinforce the importance of targeted stimulation in modulating cross-frequency interactions for task-specific memory consolidation.

Notably, a significant positive correlation between behavioral performance and EEG feature was observed exclusively in the L3 condition of the personalized TMR group. Differences in accuracy were significantly associated with SW-spindle coupling, highlighting the critical role of cross-frequency interactions in supporting memory reactivation and stabilization. SW-spindle coupling is thought to facilitate hippocampal–neocortical communication, enabling the transfer of reactivated memories from temporary hippocampal storage to stable cortical networks \cite{schreiner2023respiration, nicolas2022sigma}. In contrast, correlations involving SW power and spindle power were not statistically significant, suggesting that coupling-based neural coordination may be a more sensitive marker of successful reactivation. These findings suggest that personalized TMR enhances neural coordination by engaging complementary processes to support both memory stabilization and updating, especially under heightened cognitive demands. Furthermore, the absence of significant correlations in the All condition indicates that sleep-dependent neural reactivation disproportionately benefits complex tasks, aligning with models of task-dependent neural plasticity which propose a greater enhancement for tasks with higher cognitive demands compared to simpler tasks \cite{stickgold2009remember, bauml2014sleep}. Collectively, these results underscore the value of integrating neural metrics into personalized reactivation protocols to refine cognitive interventions.

Multivariate classification of EEG features identified distinct neural signatures among the groups, underscoring the unique contributions of SW power, spindle power, and SW-spindle coupling to memory consolidation. Enhanced classification performance after the second stimulus highlights its role in reinforcing associative links established by the first, thereby amplifying hippocampal-cortical interactions and facilitating prioritized memory reactivation and consolidation \cite{schreiner2015boosting}. The peak classification performance observed around the 3-s mark suggests that this temporal window is critical for memory stabilization, likely driven by the coordinated activity of slow oscillations and spindles. SW-spindle coupling further underscores the importance of inter-oscillatory interactions in memory consolidation, consistent with its established role in synchronizing neural networks to support efficient reactivation and integration of memory traces \cite{nicolas2022sigma, niknazar2015coupling, krugliakova2020changes}. These findings highlight the power of multivariate classification in detecting subtle differences in memory reactivation dynamics and offer robust evidence for the pivotal roles of SW and spindle dynamics in facilitating sleep-dependent memory consolidation.

This study has several limitations. First, our relatively small sample size per group may limit the generalizability of the findings; future research should include larger, more diverse populations to validate and extend these results. Second, an imbalance in the number of usable EEG trials across memory task difficulty levels—particularly the limited trial counts for L1 and L2 conditions—restricted our ability to conduct reliable electrophysiological comparisons across all difficulty levels. Although we included an All condition that pooled across levels to ensure sufficient signal quality, future studies should aim to balance cue distributions more evenly across task demands to enable finer-grained EEG analyses. Finally, the manual administration of auditory stimuli by sleep experts poses challenges for real-world applications. The development of automated, real-time EEG-based cueing systems could enhance the scalability and practicality of personalized TMR, making it more accessible as a tool for widespread cognitive enhancement.

In conclusion, our study demonstrates that personalized TMR—a personalized approach based on individual recall performance and task difficulty—enhances sleep-dependent memory consolidation. Compared to TMR and control conditions, personalized TMR led to significant improvements in memory performance under high cognitive demands, accompanied by distinct neural dynamics such as enhanced SW-spindle coupling. Multivariate classification further revealed unique neural signatures associated with personalized TMR, underscoring its potential to optimize sleep-based memory processing. These findings advance our understanding of the neural mechanisms underlying memory consolidation and offer a robust framework for developing personalized cognitive enhancement strategies with promising applications in education and cognitive rehabilitation.

\section{METHODS}
\subsection{Participants}
A total of 42 healthy participants (16 females, mean age: 26.57 ± 3.13 years) were initially recruited. All participants had normal hearing and vision and were instructed to refrain from caffeine, alcohol, and central nervous system-active drugs for 24 hours prior to the experiment. Participants were randomly assigned to one of three groups—personalized TMR, TMR, or CNT—with 14 participants per group. Due to device malfunctions, two participants from each of the personalized TMR and CNT groups were excluded. Additionally, two participants from the TMR group were excluded due to being outliers based on several variables, including sleep efficiency (SE) below 70\% and excessive wake after sleep onset (WASO) \cite{lutz2024sleep}. The final sample comprised 36 participants (12 per group; 13 females, mean age: 26.57 ± 3.13 years). The study was approved by the Korea University Institutional Review Board (KUIRB-2022-0222-04), and all participants provided written informed consent in accordance with the Declaration of Helsinki.

\subsection{Experimental design}
The experiment consisted of three main sessions: pre-sleep, sleep, and post-sleep (Fig. 1a). One week before the laboratory visit, participants completed the Pittsburgh Sleep Quality Index \cite{buysse1989pittsburgh}, Insomnia Severity Index \cite{bastien2001validation}, and Self-Rating Depression Scale \cite{zung1965self}. Upon arrival at 6 p.m., participants completed the Stanford Sleepiness Scale \cite{hoddes1972stanford} and the State-Trait Anxiety Inventory \cite{spielberger1971state}, followed by resting-state recordings and the memory task during the pre-sleep session. During the sleep session, participants were monitored with polysomnography (PSG) and slept from 10 p.m. to 6 a.m. in a controlled environment, during which auditory stimuli were presented according to the assignment of the group. Following sleep, participants took a 30-min break to reduce sleep inertia before completing the post-sleep session, which mirrored the pre-sleep session.

Participants memorized 104 semantically related word pairs (e.g., “event–festival”; Fig. 1b), presented in a randomized order (Supplementary Table 2) \cite{marshall2006boosting}. During the encoding phase in the pre-sleep session, each trial began with a 1-s fixation cross, followed by the audiovisual presentation of the word pair for 4 s. The word pair was visually presented on the screen for the entire 4-s interval, while auditory presentation occurred sequentially, with the upper word played at stimulus onset (0 s) and the lower word at 2 s. Each auditory word lasted approximately 300–900 ms, depending on its phonological length. In both the pre- and post-sleep sessions, the retrieval phase involved presenting the upper word as a cue in both visual and auditory modalities, after which participants had 10 s to type the lower word. Immediately following each retrieval trial, participants rated the difficulty of recalling the word pair using a three-level scale: L1 (“I expect little difficulty recalling this pair both immediately and after 12 h”), L2 (“I expect to recall the pair easily immediately but foresee challenges after 12 h”), and L3 (“I expect significant difficulty recalling the pair both immediately and after 12 h”). These ratings were prospective, based solely on participants' self-evaluation rather than recall accuracy, and were recorded immediately after each trial. No post hoc adjustment or reclassification of difficulty levels occurred across retrieval rounds. This prospective judgment provided a subjective measure of recall difficulty at two time points, thereby serving as an indicator of memory trace strength and helping to assess the effectiveness of the memory task and variations in retention across conditions. Additionally, during the pre-sleep session’s retrieval phase, participants received a 1-s break followed by a 4-s re-encoding period for each word pair, irrespective of their response accuracy. Tasks were conducted using Psychtoolbox 3 (http://psychtoolbox.org).

\subsection{Targeted memory reactivation}
The experimental design included three groups: a personalized TMR group, a TMR group, and a CNT group. The CNT group received no auditory stimulation during sleep and served as a baseline condition to isolate the effects of cueing strategies—specifically, the timing and repetition of cues—rather than the mere presence of sound. This design enabled a direct comparison between stimulation-based protocols and a no-stimulation baseline, focusing on how different cueing strategies influence memory consolidation. In the TMR group, a sleep expert monitored EEG signals in real-time and delivered auditory stimuli during the night when participants were stable in NREM 2 or NREM 3 for at least 10 consecutive 30-s epochs (Fig. 1c). Each auditory cue was presented for 4 s, with a 4-s inter-stimulus interval, at approximately 45 dB to ensure stimulus delivery without arousal (Fig. 1d) \cite{liu2023item}. Specifically, each cue consisted of a complete word pair presented sequentially within a 4-s window. The first word began at stimulus onset (0 s), and the second word followed approximately 2 s later. Although the inter-stimulus interval between cue onsets was fixed at 4 s, the second word typically ended around 2.3–2.9 s after onset, resulting in an effective cue-to-cue interval of approximately 5 s. This spacing ensured sufficient temporal separation between reactivation events to minimize interference \cite{cairney2018memory, antony2019sleep}. This presentation format was designed to mirror the auditory structure used during the encoding phase and to support comprehensive memory reactivation during sleep. Stimulation was paused if wake, REM, or NREM 1 stages were detected, and resumed upon return to stable NREM 2 or NREM 3. Cues from all 104 word pairs were used and presented in a randomized order during sleep, irrespective of their initial recall-difficulty levels. The total duration of stimulation averaged 77.4 ± 35.8 min, during which the TMR group received an average of 1160.3 ± 537.0 cues. These were distributed across recall levels as follows: 131.8 ± 119.9 for L1, 271.4 ± 166.3 for L2, and 757.1 ± 448.4 for L3.

To overcome the inconsistent benefits of TMR, particularly when memory performance peaks or initial learning is insufficient \cite{creery2015targeted}, we developed a personalized TMR protocol that adjusted the number of cue repetitions based on individual response accuracy and word-pair difficulty (Fig. 1d). The stimulus format—including auditory structure, timing, and sound volume—was identical to that used in the TMR group. The only difference between protocols was the frequency of cue delivery. The delivery of stimuli was categorized into three response-based conditions: 1 PRES for correct responses at L2, 2 PRES for incorrect responses at L2, and 4 PRES for any response at L3. In contrast, no stimuli were delivered for L1 items, as these were considered sufficiently consolidated during initial learning and unlikely to benefit from additional reactivation. This response-contingent protocol ensured that stimulus delivery was tailored to individual memory trace strength and learning outcomes, thereby maximizing the potential benefit of reactivation across varying levels of memory consolidation. For the personalized TMR group, stimulation lasted 67.1 ± 23.3 min on average, with 923.4 ± 318.2 cues delivered in total. Cue distribution by response-based conditions was 6.8 ± 4.1 for 1 PRES, 39.7 ± 66.4 for 2 PRES, and 877.0 ± 299.5 for 4 PRES.

\subsection{Sleep recording}
PSG recordings were conducted using the Alice 6 LDx system (Philips Respironics, Murrysville, USA). EEG signals were recorded from six scalp locations (F3, F4, C3, C4, O1, and O2) according to the international 10-20 system using Ag/AgCl electrodes. Two electrooculogram channels and two chin electromyogram channels were also recorded, with reference electrodes placed at the left and right mastoids. All signals were digitized at a sampling rate of 1,000 Hz, with data acquisition beginning only after ensuring that impedance levels were maintained below 15 k$\Omega$.

Sleep was manually scored in 30-s epochs into five sleep stages (wake, REM, NREM 1, NREM 2, and NREM 3) following the guidelines of the American Academy of Sleep Medicine criteria \cite{berry2012aasm}. Based on these stages, the sleep architecture was quantified, including total sleep time, sleep onset latency, WASO, SE, and duration and percentage of each stage of sleep.

\subsection{Behavioral data analysis}
Memory consolidation was assessed by analyzing memory accuracy and transition. Memory accuracy was quantified as the percentage of correct responses during the pre- and post-sleep sessions, with minor typographical errors disregarded to emphasize recall performance. Memory transitions were analyzed to track how memory retention, decay, and consolidation changed from pre-sleep to post-sleep sessions. These transitions were categorized into four types: CC, CI, IC, and II. Each transition type was expressed as a percentage of total responses, reflecting the dynamics of sleep-influenced memory processing across recall levels.

\subsection{EEG data pre-processing}
EEG data were processed in MATLAB 2023b with the EEGLAB toolbox \cite{delorme2004eeglab}. Data were resampled to 100 Hz and bandpass filtered between 1 and 20 Hz. Independent component analysis was applied to remove artifacts related to eye movements and muscle activity. Bad channels were identified using the \textit{pop\_rejchan} function, with kurtosis as the criterion, and interpolated using spherical interpolation. Epochs were extracted from -0.5 to 4 s relative to stimulus onset, and epochs with amplitudes exceeding $\pm$ 500 $\mu$V were excluded. For the final EEG analysis, we included only the All and L3 conditions. The number of usable trials in the L1 and L2 conditions was too small to ensure reliable averaging and valid group-level comparisons \cite{boudewyn2018many}.

\subsection{Event-related potential}
To assess brain responses to stimuli, EEG data from the personalized TMR, TMR, and CNT groups were aligned from -0.5 to 4 s relative to cue onset. Baseline correction was performed by subtracting the mean voltage from the -0.5 to 0-s interval of each trial, normalizing the data relative to pre-stimulus activity. ERP values were computed separately for recall levels (All and L3) by averaging neural responses across all channels and trials within each condition. To examine potential changes in ERP responses across repeated cue presentations, we compared the first and last cue presentations within each 4-PRES sequence in the personalized TMR group. In addition, to investigate spatial variability in ERP responses, we computed channel-wise ERPs for each of six EEG channels (F3, F4, C3, C4, O1, and O2), enabling visualization of topographical patterns across the scalp.

\subsection{Time-frequency representation}
TFR were computed for each trial using spectrograms to examine frequency-domain brain responses to stimuli \cite{tawhid2023exploring}. A 32-sample Hamming window with 24 samples of overlap and a 200-point fast Fourier transform was applied. 

The power spectrum \( P(f, t) \) was calculated as:

\begin{equation}
P(f, t) = |S(f, t)|^2,
\end{equation}

where \( S(f, t) \) represents the short-time Fourier transform of the signal, defined as:

\begin{equation}
S(f, t) = \sum_{n} x[n] \cdot w[n-t] \cdot e^{-j2\pi fn}
\end{equation}

Here, \( x[n] \) is the input signal, \( w[n] \) is the Hamming window, \( f \) is the frequency, and \( t \) is time.

The power spectrum in the 1 to 20 Hz range, which includes SW and spindle frequencies, was extracted and normalized using the baseline period. Power values were then converted to decibels. TFR data were averaged across all channels to capture neural dynamics linked to varying recall difficulty. To assess spatial heterogeneity, TFRs were computed separately for six representative EEG channels.

\subsection{Event-related phase-amplitude coupling}
To explore the interaction between SW and spindles during memory reactivation, ERPAC was analyzed using the Tensorpac Python toolbox \cite{combrisson2020tensorpac}. ERPAC was computed within the 1–4 Hz and 4–20 Hz frequency bands using the circular-linear (CL) correlation method, which quantifies phase-amplitude coupling across trials.

The CL correlation coefficient \( \rho_{cl} \) was calculated as:

\begin{equation}
\rho_{cl} = \sqrt{\frac{r_{sx}^2 + r_{cx}^2 - 2 r_{sx} r_{cx} r_{sc}}{1 - r_{sc}^2}},
\end{equation}

where \( r_{sx} = \text{corr}(\sin(\phi_t), a_t) \), \( r_{cx} = \text{corr}(\cos(\phi_t), a_t) \), and \( r_{sc} = \text{corr}(\sin(\phi_t), \cos(\phi_t)) \).

ERPAC was calculated for all channel interactions, and values were averaged across trials for each condition. To reduce trial-to-trial variability and improve robustness, ERPAC values were smoothed using a 20-sample window.

\subsection{Classification}
The multivariate classification was performed to differentiate among the personalized TMR, TMR, and CNT groups using EEG-derived features: SW power, spindle power, and SW-spindle coupling. For SW and spindle power, EEG data were extracted within their respective frequency ranges (0.5–4 Hz for SW, 12–16 Hz for spindles) and averaged across the frequency bands, resulting in feature matrices that represented the overall power within each band for each channel and participant. For SW-spindle coupling, ERPAC values were computed across the relevant frequency ranges for each channel pair and then averaged within the defined frequency bands, generating feature matrices that captured inter-channel interactions. These preprocessing steps ensured that the features preserved essential spectral and spatial characteristics while simplifying the data to facilitate robust classification and maintaining consistency across participants.

Classification was conducted over a time window from -0.5 to 4 s relative to stimulus onset to capture neural dynamics associated with auditory stimulation. Two conditions, All and L3, were analyzed to assess overall and condition-specific effects. An SVM with an RBF kernel was used due to its robustness in handling high-dimensional EEG data and ability to generalize well across datasets. Classification employed 5-fold cross-validation with 2 repetitions to ensure robust performance evaluation. Classification performance was evaluated using two metrics: accuracy and AUC. Accuracy, representing the proportion of correctly classified samples across all classes, was also calculated to provide an additional interpretable measure of classification performance. AUC was chosen as the primary metric as it measures the probability that a randomly chosen trial from one class is ranked higher than a randomly chosen trial from another class, with values ranging from 0.5 (random classification) to 1.0 (perfect discrimination). Both metrics were averaged across folds, repetitions, and trials to ensure reliable group-level comparisons. Surrogate decoding was performed by shuffling training labels 250 times to estimate chance-level performance, generating a null distribution of decoding performance under the null hypothesis of label exchangeability.

\subsection{Statistical analysis}
First, a two-way analysis of variance (ANOVA) was conducted to evaluate the impact of group and time on behavioral data. For variables showing significant main or interaction effects, pairwise comparisons were performed using Bonferroni-corrected two-sample \textit{t}-tests. Second, one-way ANOVAs were used to analyze memory performance (accuracy and transition), demographic characteristics, questionnaire, and sleep architectures, with significant findings further explored through Bonferroni-adjusted post-hoc \textit{t}-tests. Third, EEG data were analyzed using one-way ANOVAs to identify temporal or spectral differences across groups. Pearson’s correlation coefficient was then calculated to examine the relationships between behavioral and EEG results. In addition, Spearman’s rank correlation was used to assess the association between subjective difficulty ratings and memory accuracy across groups and sessions. To account for multiple comparisons in the correlation analyses, the Benjamini–Hochberg procedure was applied to control the false discovery rate \cite{benjamini1995controlling}. Finally, classification performance was assessed using a two-sided cluster-based permutation test with 1,000 permutations, comparing observed classification performance to a null distribution generated by label shuffling. Significant time points were grouped into clusters, and maxsum statistics were calculated to determine cluster-corrected \textit{p}-values via Monte Carlo simulations. The alpha level was set at 0.05.

\section*{Data availability}
The behavioral and EEG datasets analyzed during this study are available on the Open Science Framework at \url{https://osf.io/3g8rm}. 

\section*{Code availability}
The custom code used for the analyses presented in this manuscript is available on GitHub at \url{https://github.com/GihwanShin-ku/Personalized-TMR}.

\section*{Acknowledgements}
This work was partly supported by Institute of Information \& Communications Technology Planning \& Evaluation (IITP) grant funded by the Korea government (MSIT) (No. RS-2019-II190079, Artificial Intelligence Graduate School Program (Korea University)) and the National Research Foundation of Korea (NRF) grant funded by the MSIT (No. 2022-2-00975, MetaSkin: Developing Next-generation Neurohaptic Interface Technology that enables Communication and Control in Metaverse by Skin Touch).

\section*{Author contributions}
G.-H.S., Y.-S.K., and S.-W.L. designed the study. G.-H.S. and Y.-S.K. performed the experiments and collected the data. G.-H.S. and S.-W.L. developed the methodology and conducted data analysis. G.-H.S. drafted the manuscript with input from all authors. Y.-S.K., O.S., and S.-W.L. revised the manuscript critically and contributed to the important intellectual content.

\section*{Competing interests}
The authors declare no competing interests.


\bibliography{sn-bibliography}

\section*{Figure Legends}
\textbf{Figure 1.} Experimental setup. \textbf{a} The experiment comprises three main sessions: pre-sleep, sleep, and post-sleep. Participants were randomly assigned to one of three groups during the sleep session: the personalized TMR, TMR, and CNT groups. The questionnaire (Q) was administered during the wakefulness periods, alongside tasks such as word-pair learning and a 5-min eyes-closed resting state (RS) in both pre- and post-sleep sessions. \textbf{b} In the encoding phase of the pre-sleep session, participants were exposed to 104-word pairs presented via audiovisual stimuli. In the retrieval phase (pre- and post-sleep sessions), participants recalled and typed the associated word upon viewing the learned cue. Immediately following each retrieval trial, participants rated their expected difficulty in recalling the word pair both immediately and after a 12-h delay using a three-level scale: L1, L2, and L3. These prospective self-ratings were used to classify word-pair difficulty levels and were not updated or redefined across retrieval rounds. Re-encoding was provided only during the pre-sleep retrieval phase. \textbf{c} A hypnogram displaying the sleep session for one participant shows the NREM 2 and NREM 3 stages, during which the stimuli were administered, indicated by gray shading. \textbf{d} Stimulation protocols varied among groups: the personalized TMR group adjusted the number of presentations (PRES) based on response correctness—1 PRES for L2 correct responses, 2 PRES for L2 incorrect responses, and 4 PRES for L3 correct and incorrect responses. The TMR group received the same stimuli uniformly, while the CNT group received no stimulation.

\textbf{Figure 2.} Memory accuracy. Memory accuracy for all trials and for each difficulty level (L1, L2, and L3) is shown for the personalized TMR, TMR, and CNT groups. In the top row, the light-colored bars represent pre-sleep session accuracy and the dark-colored bars represent post-sleep session accuracy. The bottom row displays the difference in accuracy between the pre-sleep and post-sleep sessions. Significant differences (two-sample \textit{t}-tests with Bonferroni correction, \textit{p} $<$ 0.05) are marked with asterisks.

\textbf{Figure 3.} Memory transition. Memory transition ratios for four outcomes (correct-correct, correct-incorrect, incorrect-correct, and incorrect-incorrect) are presented for the personalized TMR, TMR, and CNT groups. The top row shows a heatmap of transition ratios, and the bottom row presents a bar plot of mean transition values and standard errors, calculated across participants and recall levels. Significant differences (two-sample \textit{t}-tests with Bonferroni correction, \textit{p} $<$ 0.05) are indicated with asterisks.

\textbf{Figure 4.} Electrophysiological responses to auditory stimuli during sleep. \textbf{a} Average event-related potentials (ERP) are displayed across all channels for the three groups in response to auditory cues. Each auditory cue consisted of a word pair presented sequentially within a 4-s window: the first word was delivered at stimulus onset (0 s), followed by the second word approximately 2 s later. The baseline period (-0.5 to 0 s) is indicated by a black bar, and stimulus onset is marked by a dashed line. Colored lines indicate group-specific ERP waveforms, and horizontal bars above the waveforms denote time intervals showing significant group differences. \textbf{b} Time-frequency representations are averaged across channels, with the top row displaying group-level spectral responses and the bottom row highlighting significant intergroup differences. \textbf{c} Event-related phase-amplitude coupling results illustrate slow wave (SW)-spindle coupling, with the top row presenting group results and the bottom row indicating significant differences. \textbf{d} Comparisons of SW power, spindle power, and SW–spindle coupling across groups reveal significant differences; significant findings from two-sample \textit{t}-tests with Bonferroni correction (p $<$ 0.05) are marked with asterisks. \textbf{e} Pearson’s correlations between behavioral and EEG results indicate significant interactions, marked with a dagger (\textit{p} $<$ 0.05, FDR correction).

\textbf{Figure 5.} Classification results. Multivariate classification performance for distinguishing personalized TMR, TMR, and CNT groups in All and level 3 (L3) conditions based on EEG-derived features (slow wave (SW) power, spindle power, and SW–spindle coupling) is presented. The shaded solid line represents observed decoding performance, while the shaded dashed line shows surrogate decoding results from 250 label shuffles. Accuracy values are presented with means and standard errors. The horizontal black solid line indicates chance-level performance, and the lower horizontal line highlights statistically significant time points, identified via a two-sided cluster-based permutation test with 1,000 randomizations.

\end{document}